\documentclass[]{IEEEapm}

\jmonth{February}
\pubyear{2023}

\begin{document}
	
	\title{Implantable and Ingestible Antenna Systems: From imagination to realization}
	
	\author{Abdul Basir, \emph{Member, IEEE}, Youngdae Cho, \emph{Grad. Student Member, IEEE}, Izaz Ali Shah, \emph{Grad. Student Member, IEEE}, Shahzeb Hayat, \emph{Grad. Student Member, IEEE}, Sana Ullah, \emph{Grad. Student Member, IEEE}, Muhammad Zada, \emph{Grad. Student Member, IEEE}, Syed Ahson Ali Shah, \emph{ Grad. Student Member, IEEE}, and Hyoungsuk Yoo, \emph{Senior Member, IEEE}}
	
	\affil{Department of Electronic Engineering, Hanyang University, Seoul 04763, South Korea. }

	\maketitle
	
	\markboth{IEEE Antennas and Propagation Magazine}{An IEEE Antennas and Propagation Society Publication}
	
	\begin{receivedinfo}%
		Abdul Basir and Youngdae Cho are co-first authors and contribute equally. This work was supported by the National Research Foundation of Korea (NRF) grant funded by the Korean government (MSIT) (No. 2022R1A2C2003726) and the Institute of Information and Communications Technology Planning and Evaluation (IITP) funded by the Korean Government Ministry of Science and ICT (MIST), under Grant 2022-0-00310. (Corresponding author: Hyoungsuk Yoo, e-mail: hsyoo@hanyang.ac.kr)
	\end{receivedinfo}
	
	\begin{abstract}
		Biomedical implantable technologies are life-saving modalities for millions of people globally because of their abilities of wireless remote monitoring,  regulating the abnormal functions of internal organs, and early detection of cognitive disorders. Enabling these devices with wireless functionalities, implantable antennas are the crucial front-end component of them. Detailed overviews of the implantable and ingestible antennas, their types, miniaturization techniques, measurement phantoms, biocompatibility issues, and materials are available in the literature. This article comprehensively reviews the design processes, design techniques and methods, types of antennas, electromagnetic (EM) simulators, and radiofrequency (RF) bands used for implantable and ingestible antennas. We briefly discussed the latest advancements in this field and extended their scope beyond conventional implantable applications. Their related issues and challenges are highlighted, and the performance enhancement techniques have been discussed in detail. All the scoped implantable applications have been covered in this review. A standard protocol has been devised to provide a simple and efficient roadmap for the design and realization of the implantable and ingestible antenna for future RF engineers and researchers. This protocol minimizes the errors in simulations and measurements by enhancing the agreement between simulated and measured results and simplifies the process of development of implantable and ingestible antennas. It generalizes the process from idea-to-realization-to-commercialization and provides an easy roadmap for the industry. 
	\end{abstract}
	
	\begin{IEEEkeywords}
		Circular polarization, diplexer, implantable and ingestible, Radio-frequency, SISO, MIMO, neural implants, wideband.
	\end{IEEEkeywords}
	
	\section{Introduction}
	
	The enormous developments in biomedical engineering and wireless technologies have advanced our knowledge of human-body electrical systems and related phenomena. These significant trends have not only revolutionized healthcare systems but also improved the lifestyle of patients by shifting in-person hospitalization to remote monitoring using implantable wireless medical devices (IWMDs). Typical health care of the IWMD applications include retinal prostheses, neural recording and stimulation, intracranial pressure measurement, capsule endoscopy, prevention of strokes through stimulation/early detection, blood sugar/glucose level monitoring, hypopnea syndrome diagnosis, cancer treatment, recovery of hearing loss, and endovascular tracking and smart sensing \cite{r1, r2, r3, r4,r5,r6,r7,r8,zada_Lens, r9, r10, stent_review, r11,r12,r13,r14,r15, r16}. The IWMD applications can be divided into four subcategories: monitoring, telemetry, stimulation, and treatment. The process of these applications starts with the collection of data by the IWMDs and transmitting this data wirelessly to on-body devices or receiving a wireless command from an outside controller to IMWD. A remote healthcare monitoring architecture through IWMDs is shown in Fig. 1.  As front-end radio-frequency (RF) components, implantable and ingestible antennas play a crucial role in controlling the operations of IWMDs owing to their inherited wireless performances. Therefore, the performance of these IWMDs depends solely on the acute performance of the implantable antennas; If the implantable antenna fails to perform, the IWMD is rendered useless, thereby risking the lives of the patients. 
	
	The design of implantable antennas starts with the selection of the frequency band. The selection of an appropriate band for real-time telemetry must consider many factors, such as the type of application, size of the implant, implantation depth, implantation site, transmission rate, range of telemetry, patient safety, and the region in which it can be used \cite{r17}.  Unlicensed bands such as the medical device radio band (MedRad: 401--406, 413--419, 426--432, 438--444, 451--457 MHz), medical implantable communication service (MICS: 402--405 MHz), or Industrial, Scientific, and Medical (ISM: 433.1--434.8, 868--868.6, 902.8--928, 2400--2438.5 MHz) bands are used for IWMDs \cite{r18}. The lower frequency band, that is the Medical Implant Communication System (MICS) band, is preferred because of lower transmission losses, longer ranges, and ensured tissue safety; however, the bandwidths are narrow and data rates are low at lower bands \cite{r19}. Unlike lower frequency bands, higher frequency bands are favorable for high data rates; however, they lead to high transmission losses, high specific absorption rate (SAR), and shortening of the transmission range. Moreover, miniaturized implantable antennas can be achieved at higher frequencies \cite{r20}. Thus, the selection of an appropriate frequency band is crucial and subjective to telemetric needs and types of applications. Deep body implants use moderate frequency bands to achieve a high data rate with miniaturized antenna geometry and an acceptable transmission range.
	\begin{figure*}[t!]
		\centering
		\includegraphics[width=35pc]{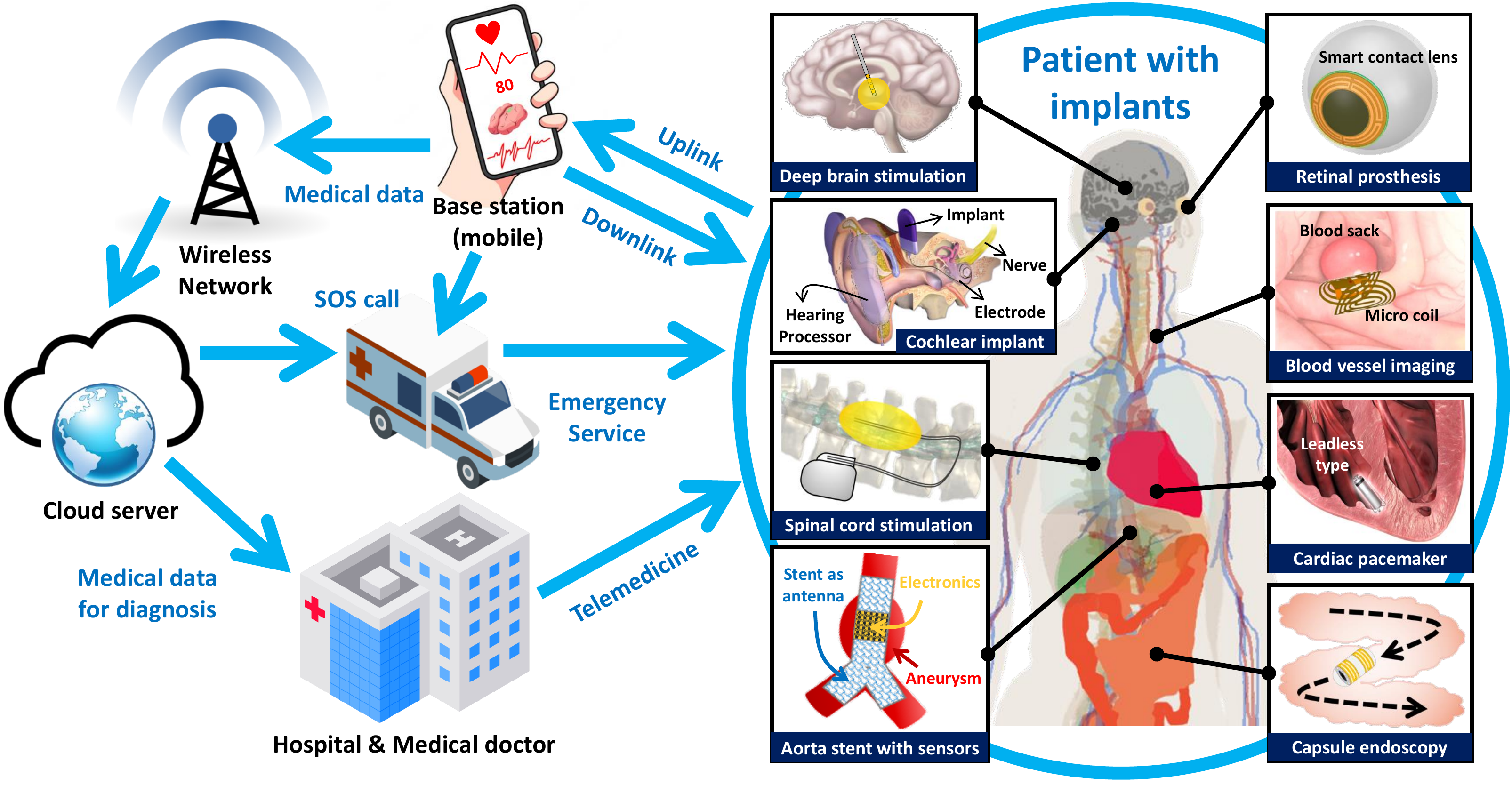}
		\centering
		\caption{A detailed overview of the implantable and ingestible devices applications for remote monitoring, architectures of the IWMDs and information flow to and from them \cite{r1, r2, r3, r4,r5,r6,r7,r8,zada_Lens, r9, r10, stent_review, r11,r12,r13,r14,r15, r16}.}
		\label{fig_1}
	\end{figure*}
	
	\begin{table}[t!]
		\caption{\label{tab:t1} Summary of the reviews of the implantable and ingestible antennas in literature and a detailed comparison.}
		\centering{\includegraphics[width=25pc]{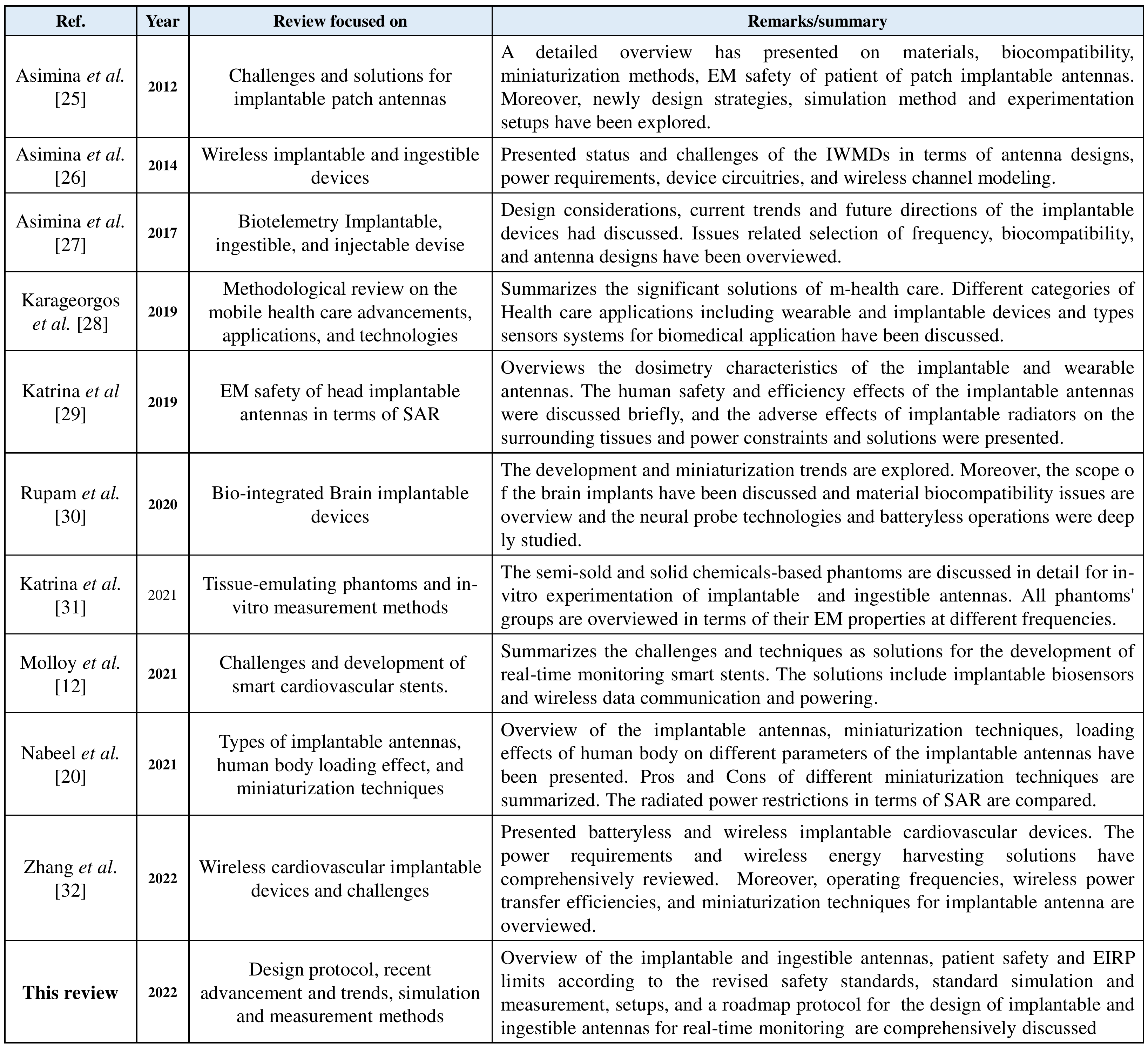}}
	\end{table}
	Over the past decade, implantable antennas have been in the research spotlight, and advancements have been made to achieve improved performance. Accurate simulations and measurement setups have been devised to standardize the design and realization of these antennas \cite{r21}. The design of an antenna within a device-like environment minimizes errors between engineering designs and real-world applications \cite{r21}. Further, exploring value-aided techniques (that is, circular polarization, flexible substrates, and metasurfaces) for implantable devices has the potential for improved performance and increased transmission range. Moreover, exploring the irregular effects of heterogeneous tissue environments of the human body on different antennas can lead to overcoming the sensitivity factors of implantable antennas \cite{r23}. An overview of the abovementioned techniques, concepts, and developments can familiarize implantable antenna designers with realizing and standardizing state-of-the-art IWMDs. Therefore, this paper presents a comprehensive review on the techniques, design processes, and recent advancements in the field of implantable and ingestible antennas. Table I presents an overview of the previous reviews on the implantable and ingestible antennas \cite{stent_review,r18,  r31,Asimina_2014,Asimina_2017,Asimina_2019,Katrina_SAR, Rupam_2020,  mat_review, Zhang_review}. It summarizes the trends and design techniques of implantable and ingestible antennas. As a comparison to start-of-the-art review articles, this review presents the recent trends, a standard design protocol, accurate simulation and measurement setups, and a roadmap for future development.   
	
	\begin{figure*}[t!]
		\centering
		\includegraphics[width=25pc]{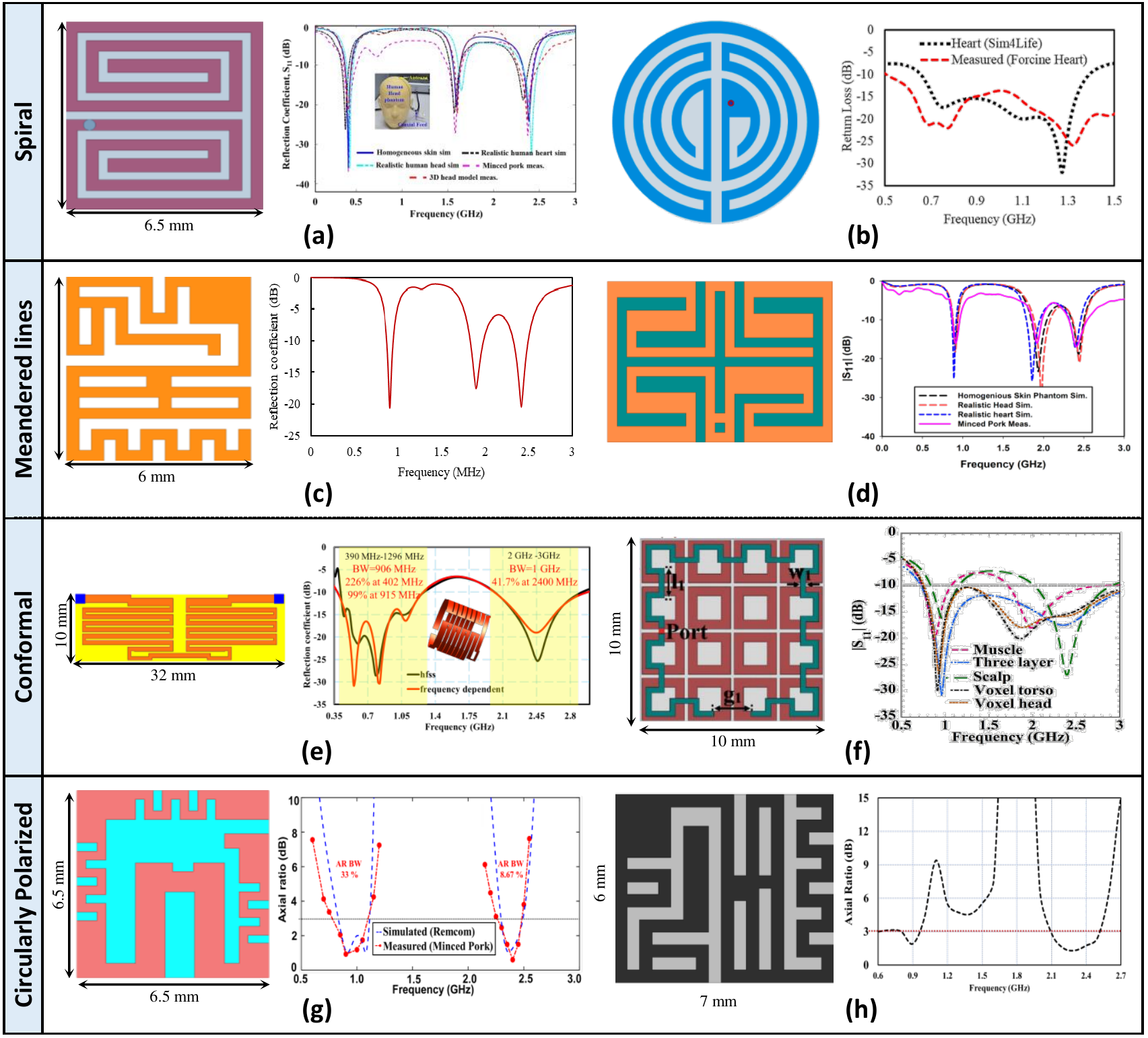}
		\centering
		\caption{Geometeries and related results of different types of implantable and ingestable antennas. (a), (b) Spiral multiband and ultrawideband implantable antennas \cite{r40,r4}. (c),(d) meandered line multiband implantable antennas and S$_{11}$ \cite{r1,r30}. (e),(f) Conformal implantable antennas for capsule endoscopes their S$_{11}$  \cite{r20, r27}. (g), (h) dual-band circular polarized planar antennas and  their axial ratios \cite{r39,r19}.}
		\label{fig_2}
	\end{figure*}
	
	\section{IMPLANTABLE AND INGESTIBLE ANTENNAS}
	In  the literature, antenna topologies including monopoles \cite{monopole_new, r25}, dipoles \cite{r17,r24, 8351709, r61}, loops \cite{r20, r27, r28}, patches, and planar inverted-F \cite{r31, r30, r29,  r32,r33, r34,r35,r36, r37,MOTL_Basir} have been employed for implantable and ingestible antennas. To fit these antennas inside a small IWMD, the implantable antenna geometries can be categorized into planar and conformal geometries \cite{r20,r40, r39, r38}. Miniaturization strategies such as spiral shapes, meandering lines, stacked-patches, the use of high-permittivity substrates and superstrates, shorting pins, and open-ended ground slots are used to achieve small sizes \cite{r1,r20,r21,r29, r32,tung2022miniaturized,r42,r41}. Planar antennas have achieved ultra-miniaturized geometries to fit inside small IWMDs, such as leadless pacemakers and brain devices in which the device diameter is less than 6 mm \cite{r40, r41, r42}. However, with such small dimensions, only the high-frequency band (2.4 GHz) can be achieved incurring increased losses, high sensitivity of the antenna, and less efficient systems. Moreover, some of these antennas have ground slots, which increase the sensitivity of these antennas to the onboard circuitries owing to coupling with the PCB of the device \cite{r2,r3,r4, r40, r39, r42, r43}. On the other hand, conformal antennas experience less coupling effect and improved gain due to the allowed large size \cite{r20, 62-new}. These antennas are wrapped inside the package of the IWMDs in such a way as to occupy less space inside the device. Conformal antennas are mostly used in endoscopes or leadless pacemakers where the devices have cylindrical geometries to provide smooth curvature for antenna wrapping. Based on miniaturization techniques, achieved bands, geometries, and enhancement techniques, some recent works have been shown in Fig.~\ref{fig_2}. Figures~\ref{fig_2}a and b show spiral triple-band and wideband antennas  presented designed and measured in \cite{r40,r4}, respectively. The meandered line tri-band antennas are optimized and tested for small devices such as leadless pacemakers in \cite{r1,r30}, as shown in Fig. \ref{fig_2}c and d. Conformal antennas with dualband circular polarization and wideband characteristics are designed for capsule endoscope in \cite{r20} and presented in Fig. \ref{fig_2}e. Authors in \cite{r27}, have designed a metamaterial loaded dualband circular polarized conformal antenna (Fig. \ref{fig_2}f) for a capsule endoscope. Figures \ref{fig_2}g and h show dualband circular polarized planar antennas for deep-body implantations \cite{r39,r19}. 
	
	\begin{figure}[t!]
		\centering
		\includegraphics[width=15pc]{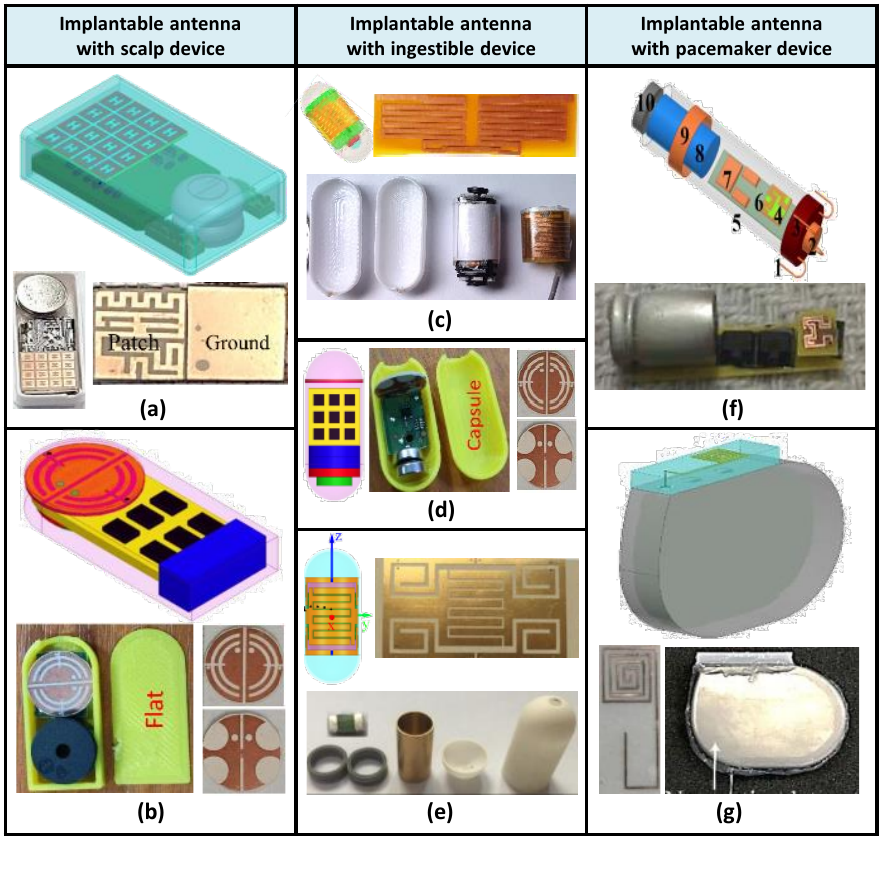}
		\centering
		\caption{The geometries of the full-package devices for the simulations and measurements of the implantable and ingestible antennas. (a) Device for brain, heart, and stomach implantable antennas \cite{r19}. (b) Antenna inside the flat-type device for deep implantations \cite{45-new}. (c), (d), (e) antennas inside the capsule endoscope \cite{r20,45-new,46-new}. (f) and (d) antennas with leadless and conventional pacemakers systems \cite{47-new, 48-new}.}
		\label{fig_3}
	\end{figure}
	
	\section{DESIGN PROCESS AND PROTOCOL}
	
	The design of the implantable antenna is completely different from that of free space antennas because of the lossy and heterogeneous operating environments \cite{r4}. The human body consists of multiple layers (bone, muscle, fat, and skin), and their dielectric properties (conductivities and dielectric constants) vary across the layers and across tissue (termed heterogeneous in biological sciences). Therefore, the operating environment of the IWMD is different and more complex than that of devices operating in free space. Owing to the heterogeneous and lossy behavior of human tissues, the development of these IWMDs is constrained by certain factors. For example; the communications range and data rates are inevitably reduced due to signal loss in the lossy tissue mediums and reflections between layers. Moreover, the varying permittivity behavior of the tissues causes impedance mismatches and frequency detuning. These issues complicate the design of antennas for these devices. Furthermore, these devices have limited available space for the device components in such a way that all the components including the antenna, must be miniaturized. However, miniaturization makes it narrow-band, less efficient, and highly sensitive to the surrounding environment. The performance of these narrow-band antennas is severely affected if the implantation site is changed or a conductor, such as a PCB or battery placed near them. In either case, the operation frequency of the antenna shifts to either the higher or lower side, causing communication failure. Therefore, it is necessary to standardize the design process of the implantable and ingestible antennas.
	
	For the accurate and stable design of implantable and ingestible antennas for real-life applications, three important factors; device architecture, simulations, and measurement setups, have been discussed in the literature. An overview of each of these is as follows. 

	\subsection{Realistic device architecture}
	
	Before 2014, all related studies used a biocompatible coating layer for insulation of the antenna to avoid direct contact with the tissue in both simulations and measurements \cite{r43}. However, in real-world implantable applications, the antenna must be placed inside a device with a complete device package. Therefore, the implantable antennas were designed prior to the work by Liu et al. \cite{r43}, and were rarely used within the real devices. In \cite{r44}, the authors designed a capsule endoscope and an antenna inside it. This study presented interesting effects of the packing on antenna performance. Following this work, in 2016, two device architectures were designed by Imran et al. \cite{r44} for skin implants. The geometrical parameters of the antenna were modified for the effects of the device. However, neither study measured the antenna within the device; therefore, the simulated and measured results were not matched.  To completely analyze the Packaging effects on the antenna,  Rupam et al. \cite{r49} designed a wideband endoscopic antenna, comprehensively  analyzed the antenna performance inside a realistic device, and set the trend for future implantable antenna engineers and researchers.
	
	Various devices have been designed based on commercially available implantable devices for application-oriented implantable antenna design and measurements. For example, Rupam et al. \cite{r49}, designed and fabricated a capsule-type device with a diameter and length of 10.25 mm and 20.5 mm, respectively, for leadless pacemaker applications. In \cite{r3}, Ahson et al. designed two device architectures for scalp implantable devices with small volumes of 344 mm$^{3}$ and 406 mm$^{3}$. Following this trend, the research group led by Prof. Yoo from South Korea and other research groups developed multiple devices including leadless pacemakers \cite{47-new,zada_ultra}, capsule and flat-type devices for deep-body implants \cite{r1, r4, r25,45-new,46-new, 48-new, r38,r49}, and capsule endoscopes \cite{r4, r20,45-new,46-new, r51,r52}. Figure. \ref{fig_3} presents a detailed overview of the devices used for simulations and measurements of implantable and ingestible antennas. 
	
	\begin{figure}[!t]
		\centering
		\includegraphics[width=30pc]{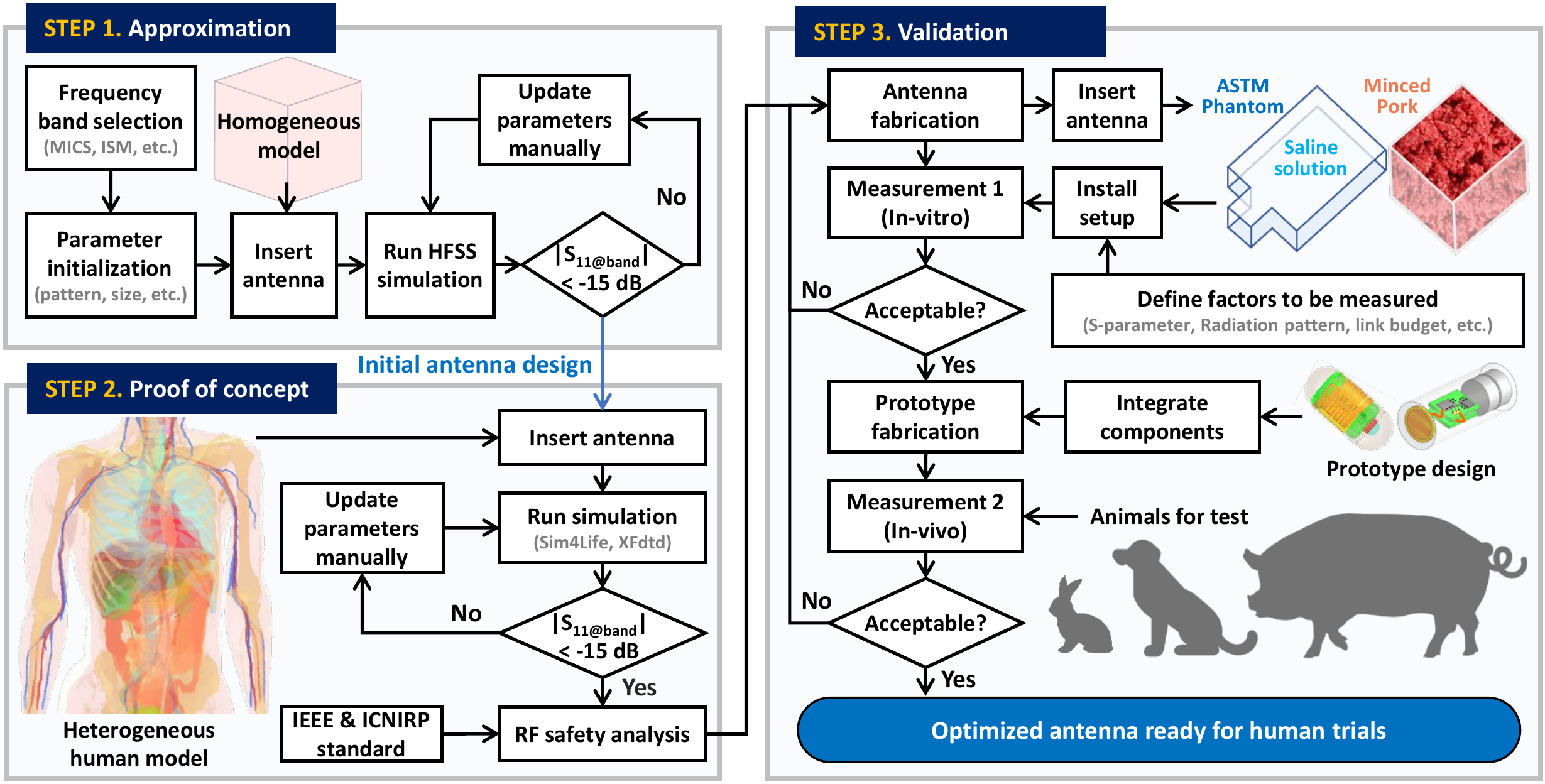}
		\centering
		\caption{A roadmap protocol for the design and realization of the implantable and ingestible antennas.}
		\label{fig_4}
	\end{figure}
	
	\subsection{Simulation Protocol and Setups}
	
	Simulations in scientific research speed up the process of device development and reduce time and costs. The studies in the literature used finite element method (FEM) and frequency- and time-domain (FDTD) simulation tools such as HFSS, CST, Xfdtd, and Sim4Life which are licensed simulators with advantages and disadvantages. For example, frequency-domain simulators such as HFSS are fast, and efficient optimization of the antenna in single-layered homogeneous phantom or simple 2-3 layered phantoms. The simulation time in HFSS depends on the number of layers in phantoms, range of analyzed frequency, size of the phantom, and geometry of the device \cite{r40,r39,r37,r38,r41,r42,r43,r44,r49,r50,r51,r52}. On other hand, time-domain-based simulation tools are highly accurate but take a long time. Therefore, a standard method is needed to be followed to reduce the optimization time, precise results, and less computational resources. For a speedy design process and easy follow-up method, we devised a customized standard protocol for developing implantable antennas, as shown in Fig. 4. This protocol leads to an accurate and simple realization of implantable antennas, which starts with the aim of the targeted application. Along with the size constraints, the frequency band was selected based on safety and data rate requirements. Then, the architecture of the device is to be modeled to mimic the device-like environment and design the antenna inside the package. Further, the selection of type and size of homogeneous phantom depends on application type and implantation depth. Farooq et al. \cite{r53}, Yousaf et al. \cite{r52}, and Nabeel et al. \cite{r18} have extensively reviewed the dimensions and types of the homogenous phantoms for implantable antennas. These dimensions range from 25 mm $\times$ 25 mm $\times$ 25 mm \cite{r42} to 200 mm $\times$ 200 mm $\times$ 200 mm \cite{r20}. For deep body antennas, the electrical properties of muscles (conductivity $\sigma$ and permittivity $\varepsilon_{r}$) have been used, and the skin properties have been used for the phantoms of skin implantable antennas. A three-layered model of muscles, fat, and skin are used in \cite{r4} and \cite{48-new}. The frequency-dependent properties of different tissues are defined and tabularized by Gabriel in \cite{r57,r58} and stored on the IT'TS website for the frequency range of 10 Hz -- 100 GHz \cite{hasgall2015database}. These properties can be assigned to the tissue phantoms in the simulation environments at either of point frequency or range \cite{r59,r60}.

	\begin{table}[t!]
		\caption{\label{tab:t1} SAR and Maximum Power Levels for Compliance with Safety Standards of various Implantable and ingestible antennas.}
		\centering{\includegraphics[width=23pc]{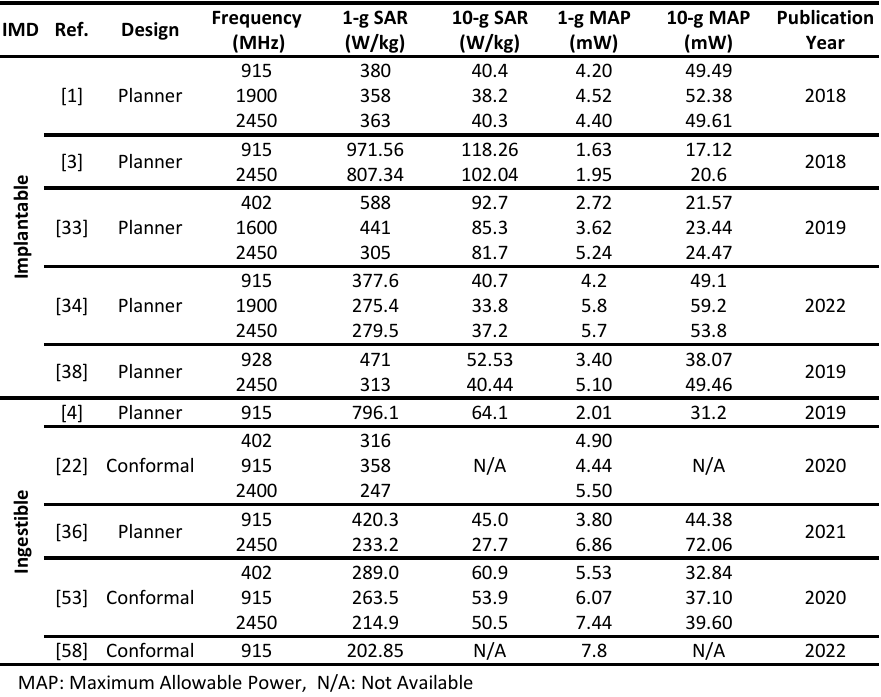}}
	\end{table}
	
	Once the implantable and ingestible antenna is fully optimized in the frequency-domain simulator, the next step is to verify its performance in the finite-domain-time-difference (FDTD) based heterogeneous environments (Sim4Life by Zurich Med Tech, CST MWS by CST Studio Suite$^{®}$, and  Xfdtd by  Remcom Inc.) \cite{r61, r62, r63}. These FDTD-based commercial simulators have realistic human models of humans, such as Duke, Ella, and Glenn in Sim4Life \cite{Izaz_1}, GUSTAV in CST, and a heterogeneous human model in Xfdtd. As shown in Step 2, the antenna performance must be verified using a realistic human model in an FDTD-based electromagnetic (EM) environment \cite{r4}. These simulation tools were used to verify the antenna performance inside different organs and assess the EM safety in terms of SAR calculations. Basir and Yoo completely verified performance and performed the SAR analysis of the capsule antenna in the Duke human model in Sim4Life  \cite{r20}. Zada et al. confirmed the simulation results of a triple-band antenna in different organs (head, colon, heart, stomach, etc.) using a realistic heterogeneous male model in Xfdtd \cite{r1}. In \cite{r42}, Yousaf et al. analyzed a multiband conformal endoscopic antenna in the GUSTAV model of CST, verified its S$_{11}$, and justified human safety in terms of the SAR. 
	
	Although antenna-integrated medical devices placed inside the body offer unique opportunities and enable significant advancements in monitoring, diagnosis, and treatment, there are concerns regarding the safety of these devices owing to their proximity to the user's body. The International Commission on Non-Ionizing Radiation Protection (ICNIRP) and the Institute of Electrical and Electronics Engineering (IEEE) have specified basic restrictions in terms of the SAR (specific absorption rate: rate of EM energy absorbed per unit mass of tissue) to prevent tissue heating due to EM radiation from the aforementioned medical devices. The ICNIRP [2010] basic restrictions confine the SAR averaged over 10 g of contiguous tissue to be less than 2 W/kg, while the IEEE C95.1-1999 standard [IEEE, 2019] limits the SAR average over any 1 g of tissue to be less than 1.6 W/kg \cite{ SAR1, SAR2}. Based on the restricted SAR limits, the effective isotropic radiated power (EIRP) of  IWMDs must be limited to ensure safety. Implantable antennas with a high EIRP cause health problems as well as interference with nearby radio devices.  An implantable antenna operating in the ISM and MedRadio bands must have an EIRP standard of -20 dB and -16 dB, respectively. Similarly, the input power of the telemetry antenna must be limited to prevent tissue damage. Moreover, the external power source should meet these standards when the implantable antenna operates in the receiver mode \cite{62-new}. A detailed SAR comparison of various implantable and ingestible antennas and their corresponding radiated power limits are presented in Table II.

	\subsection{Measurement Setups}
	
	After optimizing the implantable and ingestible antenna inside a full-package device in a homogenous model, and followed by its performance verification in an FDTD-based simulation environment in a heterogeneous human model, the antenna is ready for fabrication and measurement. The measurement process starts by connecting a flexible insulated sub-miniature-A (SMA) connector with a suitable length and diameter of  1 mm. The antenna is encapsulated in a 3D printed device (capsule- or flat-type) made of biocompatible material and implanted into the phantom for in-vitro measurement. The device must be sealed such that the liquid from the phantom does not leak into the encapsulated antenna \cite{r2, r20}. Step 3 of the protocols shows the in-vitro validation method for implantable and ingestible antennas. In the literature, different phantoms are used as measurement setups. These phantoms can be categorized as gel-based, such as saline solution \cite{r21}, semi-solid and solid chemicals-based phantoms, and dead animal tissue-based such as ground beef, minced meat, and porcine heart placed in a saline-filled American Society for Testing and Materials (ASTM) phantom. The semi-solid and solid chemical-based phantoms were comprehensively reviewed in \cite{mat_review}, and the standard saline-based and minced meat phantoms are shown in Fig. 5. A saline-filled 3D head phantom, which has the dimensions of an average young Chinese male, was used for measurements of head implanted antennas \cite{r3, r21}. A saline-filled ASTM phantom with dimensions comparable to those of the human Torso was used for the measurement of capsule endoscopes \cite{r2,r20}. Moreover, containers filled with minced pork are used for the measurements of other implantable and ingestible antennas in \cite{r20}. Because of their nearly matched dielectric properties with human tissues, minced pork phantoms provide accurate and realistic results \cite{r22}. Additionally, permittivity and conductivity can be easily controlled by adding an accurate fat ratio during the grinding process.

	After satisfying the aforementioned steps of the design protocol, the antenna is ready for use in in-vivo testing. Very few studies have reported in vivo measurements of the scattering parameters of implantable and ingestible antennas.  However, because in-vivo antennas are always used with the wireless module to perform telemetry as a standalone wireless system. The open surgery time for device implantation is limited, and the parameters of these antennas, such as the radiation pattern and S$_{11}$, etc are not required to be measured in-vivo. Instead, the matching and integration of the antenna with the device are required. For this, an approach was devised in \cite{r20, 45-new}. For example, as shown in Fig. 5h, a software-defined radio (SDR), with a standard impedance of 50 $\Omega$ and working as a transmitter, was connected to the ingestible antenna implanted in a saline-filled ASTM phantom. Furthermore, a simple monopole antenna was connected to the same type of SDR, working as a receiver, and data transmission in real-time was tested. The SDR through the implantable and ingestible antenna, placed inside the phantom, successfully communicated and sent real-time data to the SDR connected to the monopole near the ASTM phantom.

		\begin{figure}[t!]
		\centering
		\includegraphics[width=25pc]{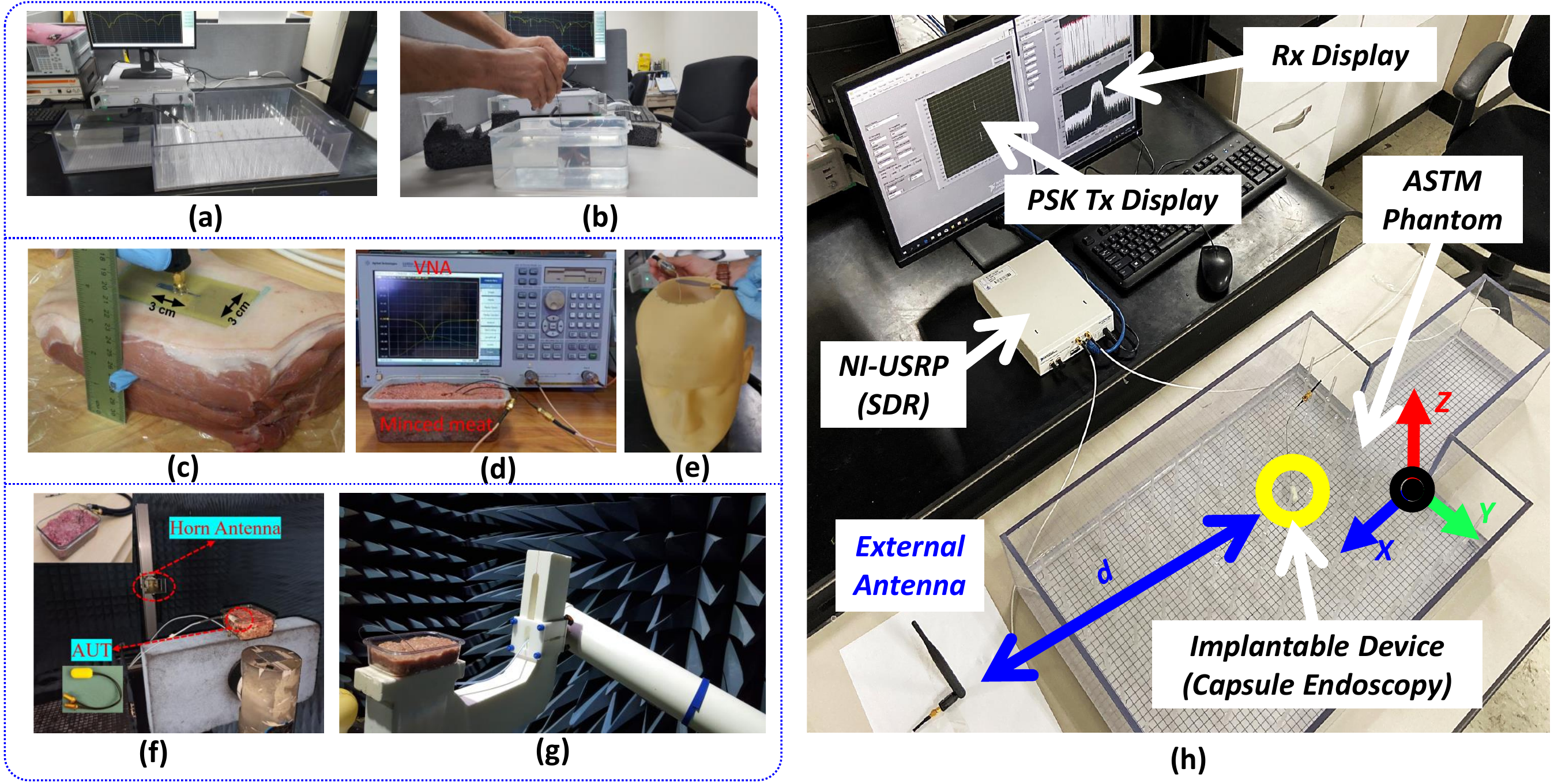}
		\centering
		\caption{Setups for S$_{11}$ and gain measurements and telemetry of the implantable and ingestible antennas. (a) A saline-filled ASTM phantom for measurement of capsule antenna \cite{r20}. (b) A saline-filled box for measurement of implantable and ingestible antennas. (c) Meat phantom for skin implantable antennas \cite{48-new}. (d) Minced-meat phantom for measurement of MIM-ingestible antennas \cite{45-new}. (e) A 3D-printed head phantom filled with saline solution for measurement of MRI-assisted implantable antenna \cite{r64}.  (f) and (g) radiation pattern measurement setups for implantable and ingestible antennas \cite{47-new} and \cite{r20}. (h) Test-bed for checking real-time telemetric capabilities \cite{r20}.}
		\label{fig_5}
	\end{figure}

	\section{RECENT ADVANCEMENTS} 
	\subsection{The Implantable and Ingestible  Antennas in Imaging Technologies}
	Magnetic resonance imaging (MRI) provides quantitative details of the imaging object, and recently, antenna designs for MRI have been proposed owing to the strong electromagnetic field generation at resonance frequencies \cite{r64,r65,youngdae_MRI,r66,r67}. Deep blood vessel imaging and structural information are crucial in fatal diseases, such as vascular plaque and intracranial brain aneurysms; however, the imaging resolution of these vessels is extremely low \cite{r67}. Implantable and ingestible antennas in MRI are termed microcoils (MCs). MCs are recommended when the patient's blood vessel images have low resolution and signal-to-noise ratio (SNR). A small catheter-based MC antenna attached to the tip of the catheter wire provides MR images of the blood vessels and surrounding tissues. An MC antenna can be implanted in the target area using a vascular access procedure.
	
	MC antennas are widely used in deep brain tissues and blood vessels. Therefore, MC antennas are designed considering the environmental effects of the deep tissues and blood vessels, such as small diameter, lossy human tissues, flowing blood, blood vessel walls, and catheter lead wire. Because the diameter of the deep vessel is very small, miniaturized and compact MC structures resonating at lower MHz bands (MRI resonance frequencies) are required for implantation in the target location. Researchers have proposed single-loop and opposite-solenoid designs integrated with catheter wire to capture the luminal images of vessels \cite{r68,r69,r70,r71}.
	
	However, a few small vessels exist in which the loop and solenoid are not practical and require ultra-compact MC antenna designs. Therefore, loopless catheter antennas, such as monopole and dipole antennas, are used to visualize small-diameter vessels \cite{r71,r72}. Although these MC designs provide higher SNR and better visualization, the dipole length and MC placement in the blood vessel remain major issues. The alternative solution and best design for inside blood vessels is the hollow-shaped MC antenna, which can be perfectly placed in the blood vessel lumen \cite{r73,r74,r75,r76}.

	Another major challenge in the MC antenna design is the blood flow inside the vessel, which creates an artifact in the MR image and increases the noise figure. The previously designed MCs, such as single-loop, solenoid, tilted, saddle-shaped, and meandering line MCs, overcome the SNR issues of deep brain tissues and inside the blood vessels; however, the blood flow blockage due to MC is not resolved in these designs. As the coil moves during intense blood flow, severe image artifacts are often observed in the arterial lumen rather than near the arterial walls \cite{r62,r76}. Therefore, an MC design with a hollow shape that is easily wrapped inside the vessel and completely covers the target area of the blood vessel is required for a large field of view, higher SNR, and homogenous field distribution inside the blood vessel. A detailed overview of the different MCs is presented in Table III.

	\subsection{Gain and Efficiency Enhancement Techniques}
	Owing to the small size and lossy tissue operation environment, the gain and efficiency of the implantable and ingestible antennas are small \cite{r19,r77}. The Friis equations show that improvement in gain strengthens the wireless communication link \cite{zada_ultra}. Therefore, only a few studies have adopted the techniques to enhance the gain and efficiency of the implantable and ingestible antennas  \cite{r19,9743651, r27}. In \cite{9743651}, the authors have used a deionized Water-Infilled cavity on the back of the scalp implantable antenna to increase its gain and efficiency.  However, works have extended applications of metamaterials to implantable antennas \cite{r19, r27}. Zada et al.\cite{r19} and Samanta et al. \cite{r27} loaded their antennas with double-positive (with positive effective permittivity and permeability) metamaterials to enhance gain and efficiency. Both works achieved at least a 1.5 dB improvement in gains and efficiencies. 
	
	\begin{table}[t!]
		\caption{\label{tab:t1} Qualitative  comparison of the implantable and ingestible micro-coil antennas.}
		\centering{\includegraphics[width=25pc]{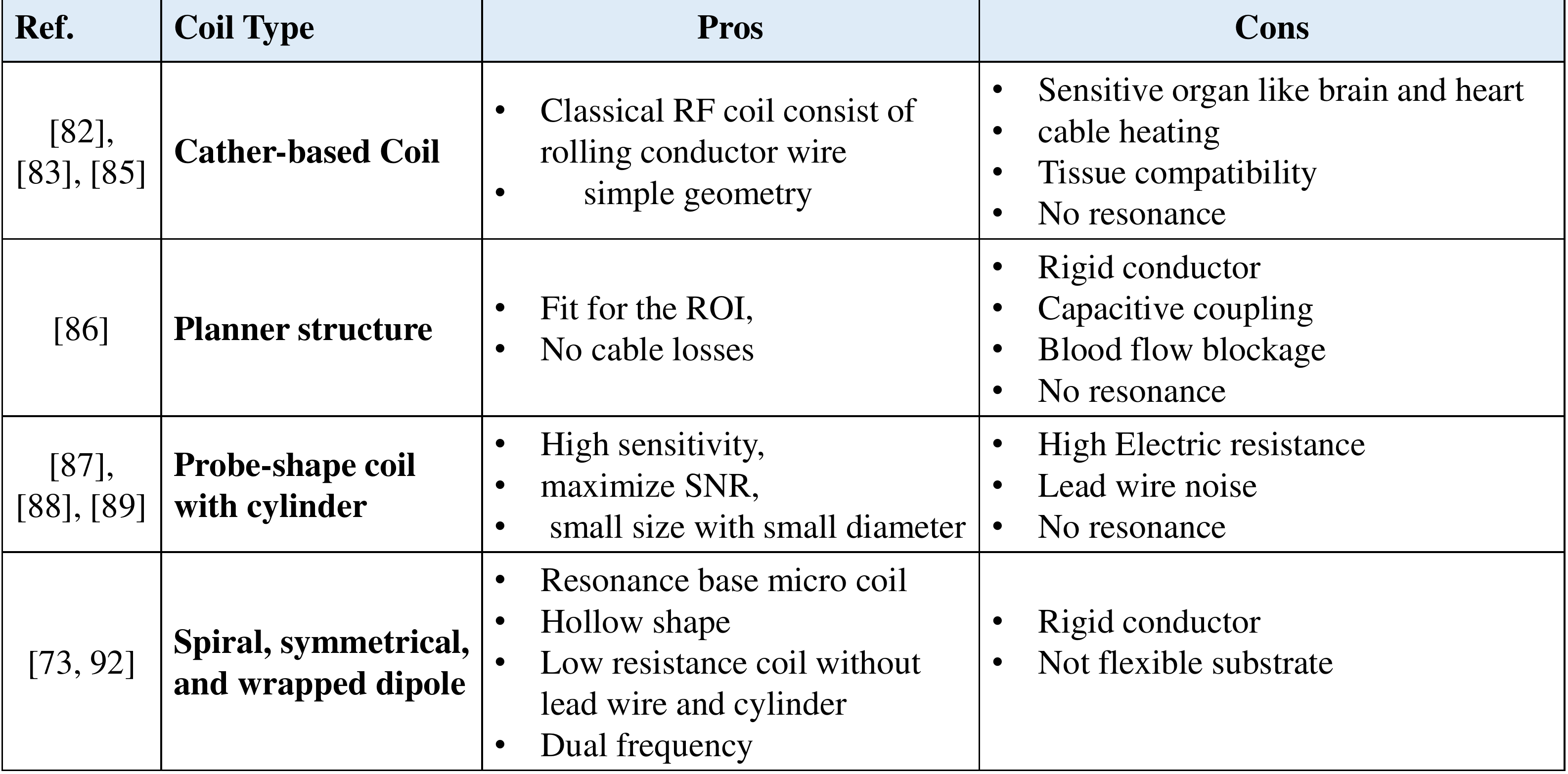}}
	\end{table}

	\subsection{Implantable MIMO Antennas}
	Due to the use of lower frequencies, where the available bandwidths are narrow, the data rates and spectral efficiency of current IWMDs integrated with single-input-single-output (SISO) antennas are low. As a solution, MIMO antennas have been introduced to improve the channel capacity and satisfy the data rate requirements of modern IWMDs.  Two-element MIMO antennas in \cite{45-new, r78,r79,r80}, and four-element MIMOs in \cite{46-new,r81,r82,r83,r84} were designed and measured for capsule endoscopic and skull implantable device applications. 
	\subsection{Implantable Multiplexer Antennas}
	Wireless multitasking, such as full-duplex telemetry, simultaneous data telemetry, and wireless battery-charging, requires extra multiplexer circuitry just after the implantable and ingestible antennas. However, these multiplexers have complex functionality and add an extra circuitry burden to the devices at the cost of making the device heavy and more power consumption.  To simplify the device design and remove the external multiplexer, a circular diplexer antenna with a diameter of 9.6 mm is designed in \cite{r51}.  The diplexer antenna has two closely-placed half-circled patches with two independent ports,  one for uplink telemetry and the other for downlink wireless power transfer. Multiplexer antennas can significantly simplify the circuitry and functions of IWMDs and extend their lives by reducing power consumption. 
	\subsection{Antennas for smart stents}
	Efficient monitoring of patients with abdominal aortic aneurysm (AAA) and sharing of biotelemetry data comprising information about certain physiological indicators are life-saving procedures \cite{stent_review}. Smart stents featuring antenna characteristics are promising candidates for endovascular aneurysmal repair and as antennas to transmit telemetric data to the external controller. Shah et al. developed a novel endovascular aortic stent system that possesses promising antenna characteristics and has the potential to be used for biotelemetry in EVAR applications \cite{r85}. Moreover, biotelemetric-enabled smart stents integrated with various sensors, including blood-flow sensors have been presented in \cite{stent_review,r86,r87, r88}.

	\section{Conclusion} 
	This paper presents an overview of implantable and ingestible antennas for device integration, as well as their related issues. Common geometries and types of antennas designed for implantable devices, simulation environments, measurement techniques, performance enhancement methods, and novel applications of implantable and ingestible antennas existing in the literature are summarized.  Furthermore, all the issues related to implantable and ingestible antennas were comprehensively discussed and important works are overviewed. Geometries of the antennas, device types, simulation and measurement setups, and techniques are categorized based on their types and applications. Each classification is discussed in terms of practicality, feasibility, and its advantages and disadvantages. A detailed study on human safety under  EM exposure for implantable and ingestible antennas is presented. Furthermore, a protocol is devised as an easy follow-up standardized method for new researchers in this field and related industries. According to this new protocol, the first step is the selection of an appropriate frequency band, designing the full package device for the antenna, and starting the optimization of the antenna inside the device. This step minimizes the gap between the design and prototyping for real-world applications. This standard protocol is a complete guideline for the design and realization of implantable and ingestible antennas as an explicit roadmap from thinking to commercialization. Moreover, performance enhancement techniques such as the gain enhancement technique, data rate enhancement (using MIMO), and simultaneous multitasking (multiplexer antennas) are discussed.
	
	\bibliographystyle{IEEEtran}
	\bibliography{Ref_all_1} 
	
	\raggedbottom
	\vfill

	\section*{Biographies}
	
	\textbf{Abdul Basir (engrobasir@gmail.com)} is a Postdoctoral Researcher with the Department of Electronic Engineering, Hanyang University, Seoul, South Korea. His research interests include implantable antennas and systems, biomedical circuits, wearable antennas, MIMO communication, metamaterial, dielectric resonator antennas, reconfigurable antennas, long-range wireless power transfer, wireless charging of biomedical implants, and stretchable electronics. \\[6pt]
	\textbf{Youngdae Cho (chb1046@hanyang.ac.kr)} is a Ph.D. candidate in the Department of Electronic Engineering, Hanyang University, Seoul, South Korea. His current research interests include implantable antennas and devices, wireless power transfer, magnetic resonance imaging, RF coils, and radio-frequency heating and safety. \\[6pt]
	\textbf{Izaz Ali Shah (izazaliuet@gmail.com)} ) is a Postdoctoral researcher in the Department of Electronic Engineering with Applied Bioelectronics Laboratory (ABL), Hanyang University, Seoul, South Korea. His current research interests include implantable antennas and devices, wireless power transfer to electric vehicles and implantable devices, MRI and RF coils, implant safety in high magnetic field systems, frequency selective surfaces, and EBGs.  \\[6pt]
	\textbf{Shahzeb Hayat (shahzebuet@gmail.com)} is a Ph.D. candidate in the Department of Electronic Engineering at the Applied Bioelectronics Laboratory, Hanyang University, Seoul, South Korea. His current research interests include implantable antenna and systems, reconfigurable antennas, MRI and RF coils, implants and tattoos safety under MRI, intravascular catheter tracking under MRI, and intracranial catheter for drug delivery.  \\[6pt]
	\textbf{Sana Ullah (sanaullah.9457841@gmail.come)} is a Postdoctoral researcher with the Department of Electronic Engineering, Hanyang University, Seoul, South Korea. His research interests include 60 GHz MIMO antennas, implantable antennas, metamaterial, RF coils for MRI, micro-coil for blood vessel imaging, wireless MRI, flexible and stretchable RF coils, and sensors with MRI receive coil. \\[6pt]
	\textbf{Muhammad Zada (muhammadzada21@gmail.com)} is a Postdoctoral Fellow at Hanyang University, South Korea, where he continues to conduct research in the field of electronic and biomedical engineering. His research interests include implantable antennas and devices, intra-oral tongue drive systems, wireless power transfer, smart contact lenses, millimeter-wave antennas, wearable sensors and antennas, frequency and pattern reconfigurable antennas, smart-textile, MRI and RF coils, microwave breast cancer detection, frequency-selective surfaces, and EBGs.  \\[6pt]
	\textbf{Syed Ahson Ali Shah (sahsonas@gmail.com)} is a Postdoctoral Researcher with the Department of Electronic Engineering, Hanyang University, Seoul, South Korea. His research interests include implantable antennas and systems, wireless power transfer to biomedical devices, implant safety, sensors integrated telemetric stents, reconfigurable antennas, and metamaterial-based antenna systems. \\[6pt]
	\textbf{Hyoungsuk Yoo (hsyoo@hanyang.ac.kr)} is a Full Professor with the Department of Biomedical Engineering and the Department of Electronic Engineering, Hanyang University, Seoul, South Korea. He has been the CEO of E2MR, a startup company, since 2017. His current research interests include electromagnetic theory, numerical methods in electromagnetics, metamaterials, antennas, implantable devices, and magnetic resonance imaging in high-magnetic field systems. 
\end{document}